# Astronomical seeing at Maidanak observatory during the year 2018

Y.A.Tillayev, A.M.Azimov and A.R.Hafizov
Ulugh Beg Astronomical Institute of the Uzbekistan Academy of Sciences
E-mail: yusuf@astrin.uz

**Abstract.** Astronomical seeing measurements were carried out at Maidanak observatory during the period from August to November 2018 using DIMM (Differential Image Motion Monitor). The median value of seeing for the entire period was determined as 0.54 arcseconds. This value was compared to the seeing data of the period 1996-2002.

The performance of the ground-based astronomical observatories is limited by the terrestrial atmosphere. Astronomical seeing is one of the key parameters characterizing overall turbulence of the atmosphere above the observatory.

Maidanak observatory is now recognized as one of the observatories in the world with the best atmospheric conditions [1]. Considering this point, it is planned to install a new modern telescope with a diameter of 4 meters in the area of the Maidanak Observatory.

Improving the optical efficiency of telescopes is one of the important tasks in the existing observatories. The optical efficiency of large telescopes depends largely on the quality of the atmospheric image and the location chosen. Astronomical seeing was last monitored at the Maidanak Observatory in 2003. No astronomical seeing measurements have been made since then. Fifteen years later (August 2018), large-scale monitoring of astronomical seeing began again. During the monitoring it was planned to identify the following:

- The current value of astronomical seeing;
- whether the astronomical seeing has changed or not;
- how much the astronomical seeing has changed;
- causes of astronomical seeing changes;
- monitor astronomical seeing on several hills at the Maidanak Observatory;
- select a suitable location for the telescope;
- increase the efficiency of the telescope.

The optical efficiency of modern large telescopes depends mainly on the astronomical seeing under the influence of the atmosphere. Astronomical seeing depends directly on the characteristics of the Earth's atmosphere. The stream of light from the star passes through the Earth's atmosphere, and the wave front coming in the same phase reaches the Earth's surface in different phases. As a result, the diffraction pattern of the star formed in the telescope vibrates. The theoretical size of a star image formed in a large telescope, excluding the Earth's atmosphere, is determined by FWHM (Full Width at Half Maximum) and is given by [2]:

$$FWHM = \frac{4}{\pi}\frac{\lambda}{D} \quad (1)$$

here D - the diameter of the telescope lens
λ - is the wavelength.

The full width of the half-maximum of intensity (FWHM) of the star profile obtained in a large telescope during a large exposure time is called astronomical seeing. Taking into account the Earth's atmosphere, the value of astronomical seeing is defined as $\varepsilon_{FWHM}$ and is determined according to the following formula:

$$\varepsilon_{FWHM} = 0.98\frac{\lambda}{r_0} \quad (2)$$

here $r_0$- Fried parameters. It is usually measured in centimeters.

There are several ways to measure astronomical seeing today. One of the most effective methods is to measure with a DIMM (Differential Image Motion Monitor) instrument [2]. The main optical mirror of the DIMM has a diameter of 279 mm and a focal length of 2800 mm. The head of the optical tube is fitted with a diaphragm and covered with a Schmidt corrector. The diaphragm has two 8 cm round holes with a distance of 20 cm between the centers. One of the slits is mounted with a 195 arcsecond wedge prism.

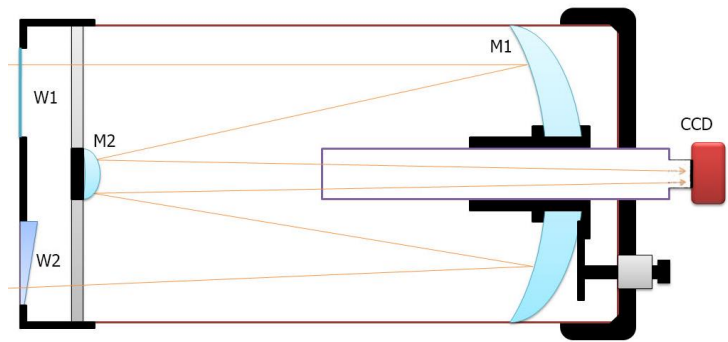

**Figure 1.** $M_1$- main optical mirror, $M_2$- secondary hyperbolic mirror, $W_1$- first slit, $W_2$- second prismatic slit

A prism mounted in the second aperture changes the direction of the light beam, and two images of the star are formed in the focus of the telescope. These images are recorded by a CCD camera. The resulting image has a vibrating effect. The function of the DIMM is to determine the displacement of the two image centers along the x and y axes relative to each other, i.e., the vibration of the star. This determines the value of astronomical seeing in units of arc seconds. The smaller the vibrational effect of the star (the smaller the astronomical seeing), the more suitable the atmosphere of the observatory. It turns out that for astronomical observations, the value of the astronomical seeing at the observation site should be small. The DIMM device was taken to the Maidanak Observatory and mounted on a 6-meter platform in early August this year. Monitoring of astronomical seeing using the DIMM telescope began on August 3 and continues to this day. According to the observations, standard stars are observed mainly at the zenith (within 30° of z distance). This is because the airmass is the smallest near the zenith. A particular star can be observed for a maximum of 3 hours. If observed for more than 3 hours, the star leaves the zenith circle of 30 °. The list of stars selected for observation is given in the Table 1. This table lists the names, magnitudes, and equatorial coordinates of some of the bright stars that were observed.

|   | Star | Apparent magnitude (m) | Right ascension ($\alpha$) | Declination ($\delta$) |
|---|------|------------------------|----------------------------|------------------------|
| 1 | α And | 2,05 | $0^h\ 08^m\ 23^s.25$ | +29° 05' 25".55 |
| 2 | β Peg | 2,48 | $23^h 03^m\ 46^s.45$ | +28° 4' 58".025 |
| 3 | α Cyg | 1,33 | $20^h 41^m 25^s.91$ | +45°16' 49".21 |
| 4 | α Lyr | 0,09 | $18^h 36^m 56^s.33$ | +38°47' 01".29 |
| 5 | α Ari | 2,00 | $02^h 07^m 10^s.29$ | +23°27' 45".94 |

**Table 1.** A partial list of the stars observed.

One of the most important aspects to consider when measuring astronomical seeing is choosing the optimal exposure time. The star should have enough time to record the vibration and this time should be as short as possible. Usually the exposure time is two and the first value is half of the second value.

In observations made using the split tool, $\epsilon_1$ va $\epsilon_2$ the length and width measurements and the vibration of the star image in the x and y coordinates display can be equated to square meters. Exposure times are set in the sequence $\tau_1=10$ and $\tau_2=20$ ms. For each exposure time, quality indicators are found and the performance on them is specified, $\tau=0$ quality indicators are determined. This value is represented by the following formula [3]:

$$\epsilon_0 = 0.5(c_1\epsilon_1 + c_1^{7/3}\epsilon_2) \tag{3}$$

Based on the observations at the Maidanak Observatory, the values of astronomical seeing obtained from 3 August to 20 November were processed and analyzed.

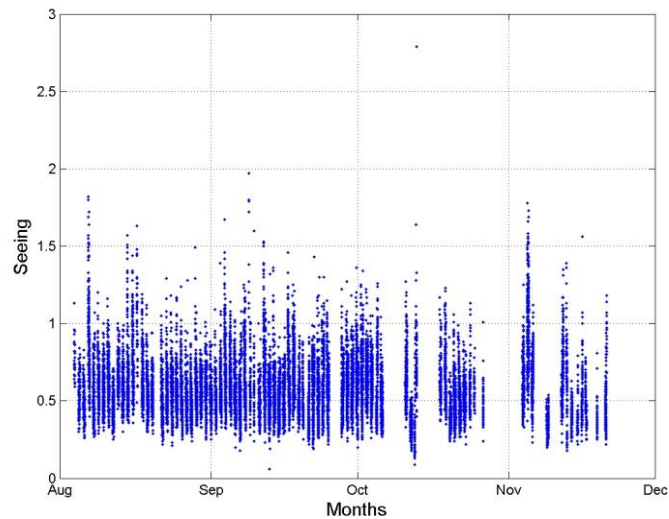

**Figure 2.** Graph of astronomical seeing measured in August-November, 2018

Figure 2 shows a graph of astronomical seeing results obtained at the Maidanak Observatory. The graph shows the number of points obtained on the astronomical seeing values during the x observation period, and the value of the astronomical seeing on the y axis in arcsecond units.

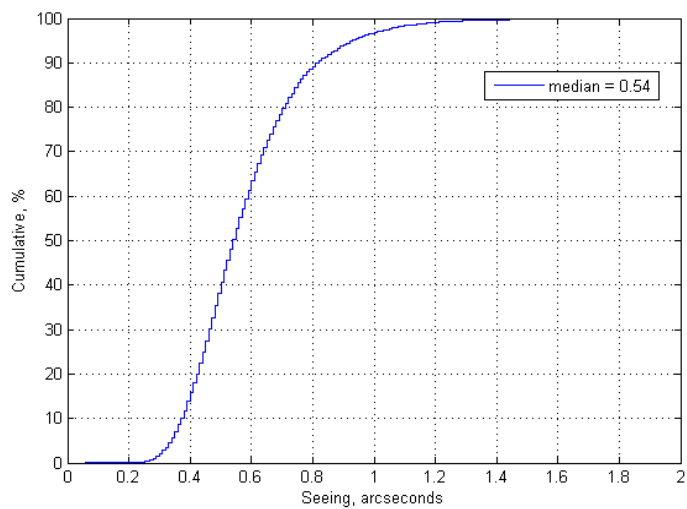

**Figure 3.** Graph of cumulative distribution of astronomical seeing for the period August-November, 2018.

The cumulative distribution of astronomical seeing is shown in Figure 3. As can be seen from this graph, the median value of astronomical seeing for about 4 months is 0.54 arcseconds.

| Month | Seeing ($\varepsilon_{FWHM}$) | | | |
|---|---|---|---|---|
| | 25% | 50% | 75% | Average |
| August | 0.46 | 0.55 | 0.67 | 0.58 |
| September | 0.43 | 0.54 | 0.68 | 0.57 |
| October | 0.42 | 0.52 | 0.65 | 0.55 |
| November | 0.39 | 0.53 | 0.74 | 0.59 |
| Median | 0.43 | 0.54 | 0.68 | 0.57 |

**Table 2.** Astronomical seeing statistics for each month for Maidanak Observatory

This table shows the average, median values for each month of astronomical seeing results, as well as values at 25% and 75%. The median value of astronomical seeing is 0.55 for August, 0.54 for September, 0.52 for October, and 0.53 for November. The value at 25% of all points obtained is 0.43 arc seconds. The value at 75% is 0.68 arc seconds. To find out how good or bad these astronomical seeing indicators are and how much they have changed, we compare the astronomical seeing results obtained using the same DIMM instrument in 1996-2002.

| Month | Number of nights | Number of measurements | $\varepsilon_{FWHM}$, arcseconds |
|---|---|---|---|
| January | 68 | 6069 | 0.79 |
| February | 66 | 7052 | 0.77 |
| March | 41 | 3679 | 0.71 |
| April | 49 | 3986 | 0.78 |
| May | 82 | 7160 | 0.70 |
| June | 93 | 7522 | 0.71 |
| July | 139 | 10948 | 0.73 |
| August | 148 | 14149 | 0.72 |
| September | 168 | 19807 | 0.69 |
| October | 135 | 15576 | 0.68 |
| November | 75 | 8061 | 0.65 |
| December | 70 | 6704 | 0.71 |
| **Total** | **1134** | **110713** | **0.71** |

**Table 3.** Median values for each month of astronomical seeing obtained at the Maidanak Observatory from 1996 to 2002 [4].

From August 1996 to October 2002, astronomical seeing results were obtained using the DIMM instrument at the Maidanak Observatory. The results obtained are presented in Table 3. This table shows the median value of each month of astronomical seeing obtained between 1996 and 2002, as well as the number of nights observed and the number of dots obtained. Accordingly, the best value for astronomical seeing was in November (0.65 arcseconds) and the worst was in January (0.79 arcseconds). Overall, the median value of astronomical seeing at the Maidanak Observatory at that time was 0.71 arcseconds. By August 2018, the Maidanak Observatory again started monitoring astronomical seeing using a DIMM instrument. The current results of astronomical seeing were compared with the results for the corresponding months of 1996-2002. The results show that astronomical seeing improved by 0.17 arcseconds in August, 0.15 arcseconds in September, 0.16 arcseconds in October, and 0.12 arcseconds in November. Overall, the astronomical seeing improved by 0.17 arcseconds.

The values of current astronomical seeing are slightly better than those measured in 1996-2002. This will further increase the effectiveness of night surveillance. However, due to the short duration of the results obtained in 2018, it is not possible to draw a complete conclusion. This requires a follow-up period of at least 1 full year or more. Long-term observations and calibrations are now planned with other instruments as well. The results obtained with different instruments and in the long run are compared and analyzed. Only then can a final conclusion be drawn.